\documentclass[referee,sn-nature]{sn-jnl}

\usepackage{graphicx}%
\usepackage{multirow}%
\usepackage{amsmath,amssymb,amsfonts}%
\usepackage{amsthm}%
\usepackage{mathrsfs}%
\usepackage[title]{appendix}%
\usepackage{xcolor}%
\usepackage{textcomp}%
\usepackage{manyfoot}%
\usepackage{booktabs}%
\usepackage{algorithm}%
\usepackage{algorithmicx}%
\usepackage{algpseudocode}%
\usepackage{listings}%
\usepackage{siunitx}

\usepackage{geometry}
\usepackage{nicefrac}

\theoremstyle{thmstyletwo}%

\theoremstyle{thmstylethree}%

\unnumbered

\begin{document}

\title[Vervelaki et al.]{Visualizing thickness-dependent magnetic textures in few-layer $\text{Cr}_2\text{Ge}_2\text{Te}_6$}


\author[1]{\fnm{Andriani} \sur{Vervelaki}}\email{andriani.vervelaki@unibas.ch}

\author[1]{\fnm{Kousik} \sur{Bagani}}\email{kousik.bagani@unibas.ch}

\author[1]{\fnm{Daniel} \sur{Jetter}}\email{daniel.jetter@unibas.ch}

\author[2]{\fnm{Manh-Ha} \sur{Doan}}\email{mando@dtu.dk}

\author[2]{\fnm{Tuan K.} \sur{Chau}}\email{tukhc@dtu.dk}

\author[1]{\fnm{Boris} \sur{Gross}}\email{borisandre.gross@unibas.ch}

\author[3]{\fnm{Dennis} \sur{Christensen}}\email{dechr@dtu.dk}

\author[2]{\fnm{Peter} \sur{Bøggild}}\email{pbog@dtu.dk}

\author*[1,4]{\fnm{Martino} \sur{Poggio}}\email{martino.poggio@unibas.ch}

\affil*[1]{\orgdiv{Department of Physics}, \orgname{University of Basel}, \orgaddress{\postcode{4056}, \city{Basel}, \country{Switzerland}}}

\affil[2]{\orgdiv{Department of Physics}, \orgname{Technical University of Denmark}, \postcode{2800}, \city{Kongens Lyngby}, \country{Denmark}}

\affil[3]{\orgdiv{Department of Energy Conversion and Storage}, \orgname{Technical University of Denmark}, \postcode{2800}, \city{Kongens Lyngby}, \country{Denmark}}

\affil[4]{\orgdiv{Swiss Nanoscience Institute}, \orgname{University of Basel}, \orgaddress{\postcode{4056}, \city{Basel}, \country{Switzerland}}}


\abstract{
Magnetic ordering in two-dimensional (2D) materials has recently emerged as a promising platform for data storage, computing, and sensing. To advance these developments, it is vital to gain a detailed understanding of how the magnetic order evolves on the nanometer-scale as a function of the number of atomic layers and applied magnetic field. Here, we image few-layer $\text{Cr}_2\text{Ge}_2\text{Te}_6$ using a combined scanning superconducting quantum interference device and atomic force microscopy probe. Maps of the material's stray magnetic field as a function of applied magnetic field reveal its magnetization per layer as well as the thickness-dependent magnetic texture. Using a micromagnetic model, we correlate measured stray-field patterns with the underlying magnetization configurations, including labyrinth domains and skyrmionic bubbles. Comparison between real-space images and simulations demonstrates that the layer dependence of the material's magnetic texture is a result of the thickness-dependent balance between crystalline and shape anisotropy. These findings represent an important step towards 2D spintronic devices with engineered spin configurations and controlled dependence on external magnetic fields.}




\maketitle

\section{Introduction}\label{sec1}

The recent discovery of long-range ferromagnetic ordering in bilayer $\text{Cr}_2\text{Ge}_2\text{Te}_6$ (CGT) and monolayer $\text{CrI}_3$ has opened the fascinating new area of two-dimensional (2D) magnetic materials~\cite{gong_discovery_2017, huang_layer-dependent_2017, huang_emergent_2020}. These materials often have different magnetic properties from those of their bulk counterparts and these properties can be strongly dependent on the number of layers. For instance, although bulk CrI$_3$ is ferromagnetic, it is antiferromagnetic in the few-layer limit~\cite{niu_coexistence_2020, liu_thickness-dependent_2019, thiel_probing_2019}. 
Furthermore, the inter-layer exchange interaction can lead to antiferromagnetic or ferromagnetic behavior depending on whether the layers are stacked in a monoclinic or rhombohedral manner~\cite{xu_coexisting_2022, tong_skyrmions_2018, song_direct_2021}. Indeed, the possibility of engineering van der Waals (vdW) heterostructures from these systems -- atomic layer by atomic layer -- affords unprecedented control over their magnetism. Moreover, unlike bulk magnets, 2D magnets are highly susceptible to external stimuli, including electric fields, magnetic fields, and strain~\cite{wang_magnetic_2022}. For all of these reasons, 2D magnets have potential applications ranging from high-density data storage and efficient information processing to sensing.  
 
 Among these materials, the ferromagnetic semiconductor CGT is especially promising for spintronic and memory storage applications. Both current~\cite{mogi_current-induced_2021} and laser~\cite{khela_laser-induced_2023} induced magnetization switching has been demonstrated in this material. It exhibits spin-orbit torque switching under very low current densities, making it promising for low-power memory devices~\cite{ostwal_efficient_2020}. In addition, CGT is a candidate for pressure-sensitive spintronic applications, because of its strong anisotropic spin-lattice coupling~\cite{chen_anisotropic_2022}. Although CGT has a low Curie temperature  $T_\text{c} = 68$~K in the bulk, which drops to 30~K in the bilayer limit~\cite{gong_discovery_2017}, several methods have succeeded at increasing $T_\text{c}$, including via doping~\cite{verzhbitskiy_controlling_2020, zhuo_manipulating_2021} and strain~\cite{oneill_enhanced_2023, siskins_nanomechanical_2022}. In particular, the application of 2.3\% strain has been shown to raise $T_\text{c}$ up to room temperature.

Because of these promising properties, its magnetic behavior and -- specifically -- the dependence of this behavior on the number of layers needs to be understood. Since the first magnetic imaging of CGT, studies have been carried out on samples of different thicknesses~\cite{noah_interior_2022, lohmann_probing_2019,noah_nano-patterned_2023,gupta_manipulation_2020, han_topological_2019}.
Few-layer CGT with thicknesses between 2 and 5~nm shows soft ferromagnetic behavior with an out-of-plane magnetic easy-axis and no magnetic domains~\cite{gong_discovery_2017}. 
In contrast, hysteretic behavior with multi-domain magnetic textures is observed in CGT flakes with thicknesses in the tens of nanometers~\cite{noah_interior_2022, lohmann_probing_2019, noah_nano-patterned_2023}. For samples thicker than about 100 nm, the presence of Bloch-type magnetic stripe domains and skyrmionic bubbles has been reported~\cite{han_topological_2019}. 
Despite these observations, the evolution of magnetic order with thickness, i.e.\ the transition from single-domain soft ferromagnetic behavior in few-layers to multi-domain structure in the thicker limit has not been investigated. Also, the type of magnetization textures forming the domains observed in the intermediate thickness range remains unclear.

In this work, we investigate the dependence of CGT's magnetic behavior on thickness down to the few-layer limit. 
We image the material's stray magnetic field, determine the corresponding magnetization configuration, and map its evolution as a function of applied magnetic field. 
Nanometer-scale magnetic imaging is carried out via a superconducting quantum interference device (SQUID) integrated on a cantilever scanning probe at 4.2~K~\cite{wyss_magnetic_2022}. 
The SQUID-on-lever's (SOL) ability to simultaneously image both the sample's topography and stray field allows us to correlate material thickness with magnetic configuration. 
Furthermore, a comparison of the measured stray-field maps with micromagnetic simulations sheds light on the form of the underlying magnetization configurations and the magnetic interactions which produce them. 

\section{Magnetization per layer} \label{sec2}

\begin{figure}[h]%
\centering
\includegraphics[width=1\textwidth]{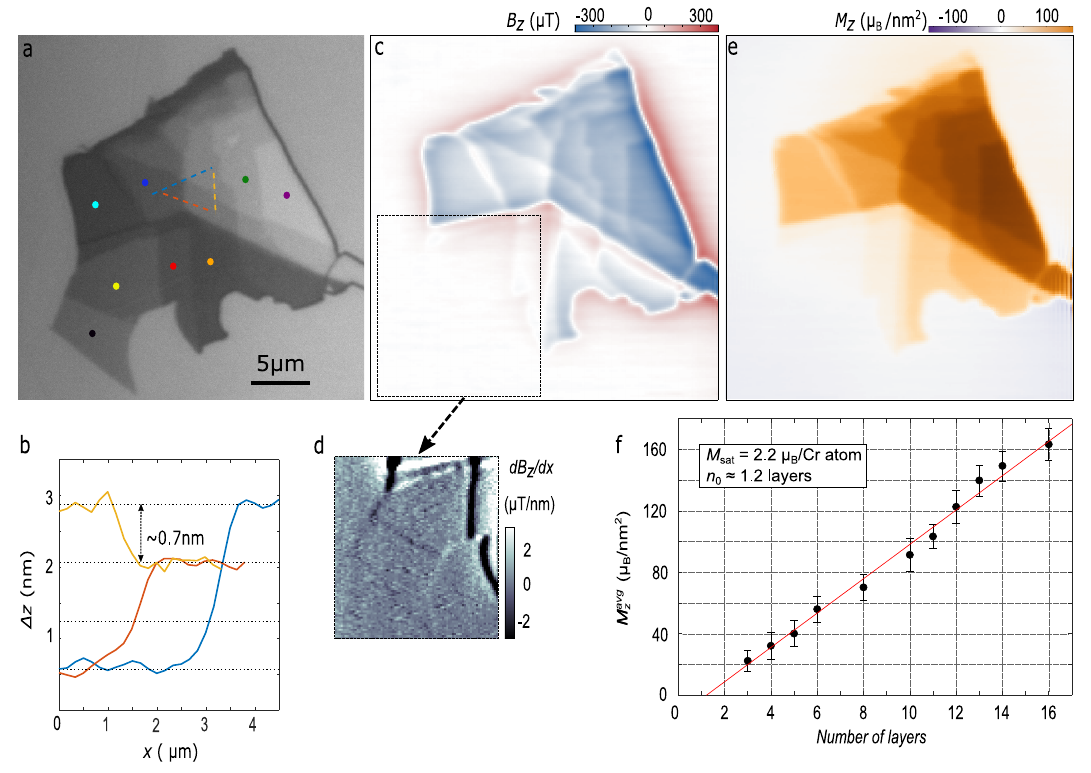}
\caption{ Magnetization per layer. (a) Optical image of the CGT flake. Regions of different thickness can be identified from the optical contrast and are indicated with black, yellow, red, orange, aquamarine, blue, green, and purple circles for thicknesses of 2, 3, 4, 6, 8, 10, 14, and 16 layers, respectively. (b) AFM scans between regions of uniform thickness showing steps of integer layer thickness along the correspondingly colored dashed lines in (a). (c) $B_z(x,y)$ measured above the sample in an applied out-of-plane field, $\mu_0H_z=-133\,\text{mT}$.  (d)  Numerical derivative along the x-axis, $\frac{dB_z(x,y)}{dx}$ of the region indicated by the dotted square in (c). (e) Out-of-plane magnetization $M_z(x,y)$ calculated from (c).  (f) $M^{avg}_z$ plotted as a function of the number of layers and its corresponding linear fit. 
}\label{fig1}
\end{figure}

CGT flakes are mechanically exfoliated on a Si/SiO$_2$ substrate. Because CGT is prone to degrade in ambient conditions~\cite{gong_discovery_2017}, flakes are covered with 10-nm-thick hexagonal boron nitride (hBN) immediately after exfoliation. Figure~\ref{fig1}a shows an optical image of a flake with regions of various thicknesses. These regions can be distinguished by their different optical contrast. Their thickness is measured via atomic force microscopy (AFM) (supplementary Figure S1), acquired by the SOL scanning probe in non-contact mode.
In Figure~\ref{fig1}b, AFM cross-sections taken across the boundaries between regions of different thicknesses show steps in integer multiples of 0.7 nm, which is the thickness of a single atomic layer. The number of layers in each region is then determined using this thickness per layer and a thickness of 1.1 nm for the first layer~\cite{gong_discovery_2017}. The estimated thickness of the flake ranges from 2 to 16 layers with integer layer steps between adjacent areas. The availability of different thicknesses on the same flake allows us to study the evolution of its magnetic behavior as a function of the number of layers.

The magnetic properties of a CGT flake are investigated by imaging the out-of-plane component of the sample's magnetic stray field, $B_z(x,y)$, in a plane above the sample. 
Initially, we apply an out-of-plane magnetic field $\mu_0 H_z=-133\,\text{mT}$ in order to saturate the magnetization of the flake. 
The resulting $B_z(x,y)$, shown in Figure~\ref{fig1}c, is peaked at the edges of the flake or at boundaries between regions of different thicknesses, as expected for a sample uniformly magnetized in the out-of-plane direction. 
The thinnest regions on the bottom-left corner of the image (indicated with black and yellow dots) are 2 and 3 layers thick and show very weak magnetic signatures, visible in $\frac{dB_z(x,y)}{dx}$, shown in Figure~\ref{fig1}d.


From $B_z(x,y)$, shown in Figure~\ref{fig1}c, we then determine the sample's out-of-plane magnetization $M_z(x,y)$, shown in Figure~\ref{fig1}e. 
By assuming that the magnetization is fully saturated along the $z$-direction and confined to a 2D plane (the sample is much thinner than the probe-sample distance), we use a reverse propagation method to solve for $M_z(x,y) = \mathbb{A} ^{-1} B_z(x,y)$ where $\mathbb{A}$ is the transfer matrix~\cite{broadway_improved_2020}. 
Figure~\ref{fig1}f shows the reconstructed magnetization averaged over areas of uniform thickness, $M_z^{avg}$, as a function of the number of layers. 
In particular, the magnetization of each area increases linearly with the number of layers.  From the slope of the linear fit, we obtain a magnetization per layer $M_z^\text{layer}= (10.9 \pm 0.8) \,\mu _\text{B}/\text{nm}^2$, equivalent to a saturation magnetization $M_\text{sat}= (2.2 \pm 0.2)\, \mu _\text{B}/\text{Cr}$ atom, which is consistent with previous reports ~\cite{suzuki_magnetic_2022, zhang_magnetic_2016, ji_ferromagnetic_2013}.  
From the non-zero horizontal intercept of the fit, we find that the flake contains $n_0= 1.2 \pm 0.6$ magnetically inactive layers. This effect is likely due to the degradation of the outer surfaces of the flake during the short exposure to air before encapsulation. Hereafter, the number of layers of CGT refers to the number of magnetically active layers, which are roughly one less than the number of physical layers. 




\section{Thickness-dependent magnetic hysteresis }\label{sec3}

To investigate the microscopic magnetization configurations involved in magnetic reversal and their dependence on thickness, we map $B_z(x,y)$ as a function of applied magnetic field. 
Figures 2a-i show the evolution of the flake's magnetization reversal with respect to out-of-plane applied field $H_z$. 
\begin{figure}[h]%
\centering
\includegraphics[width=0.84\textwidth]{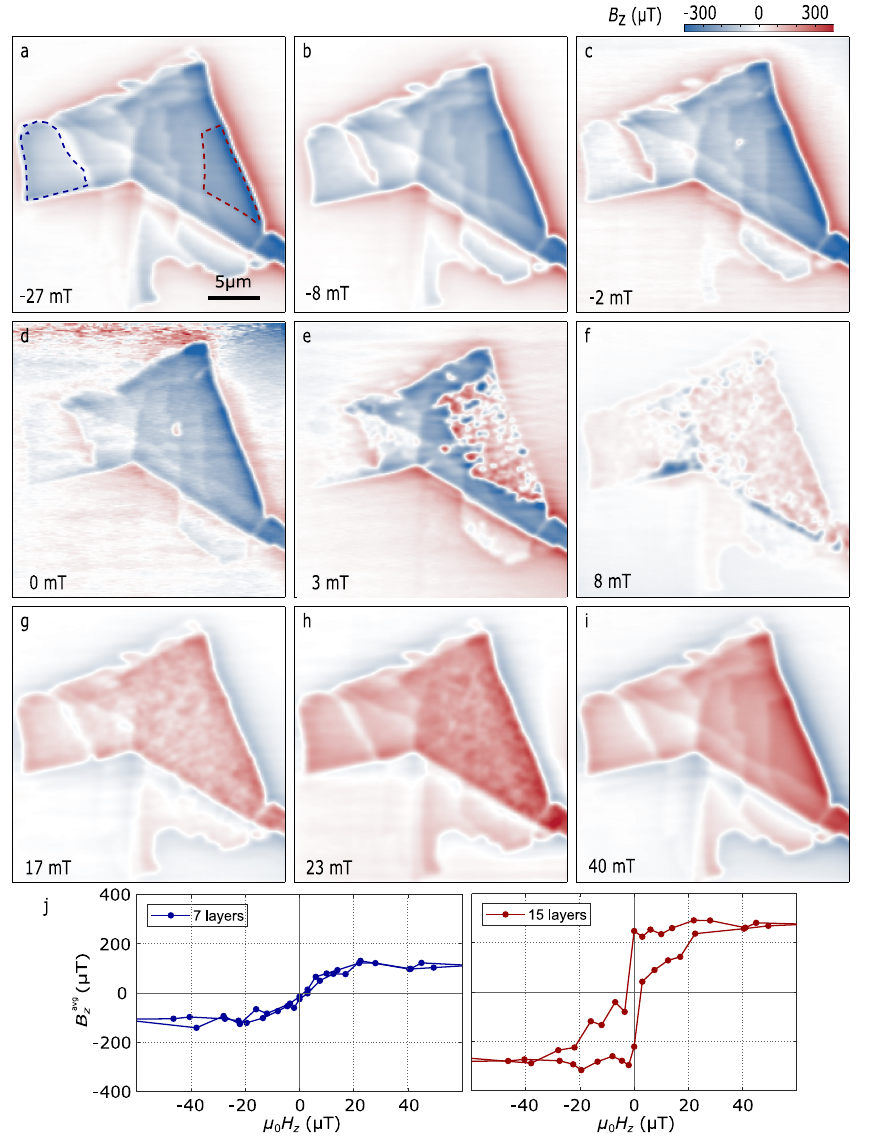}
\caption{Visualizing the out-of-plane magnetic hysteresis. (a-i) $B_z(x,y)$ measured above the sample at different $\mu_0H_z$ as indicated by the value shown in each image. $\mu_0H_z$ is slowly stepped from -133~mT to 140~mT. Blue and red dashed lines highlight the 7 and 15-layer-thick regions of the flake. Images for more values of $\mu_0H_z$ can be found in supplementary Figure~S3. (j) $B_z^\text{avg}$ for areas of constant thickness highlighted in (a) plotted against $\mu_0H_z$.  }\label{fig2}
\end{figure}
Starting in the saturated state, initialized at $\mu_0H_z=-133$~mT, the applied field is gradually stepped towards zero, into reverse field, and past saturation to 140~mT. Until zero applied field, the magnetization in the thicker regions (10-15 layers) remains almost unchanged, whereas, in the thinner parts of the flake (5-7 layers), domains start to form around $\mu_0 H_z =-10$~mT (more apparent in supplementary Figure S4).
Upon reversing the direction of $H_z$, magnetic domains also nucleate in the thicker regions of the flake. 
With increasing reverse field $H_z$, these domains spread over the whole flake and result in a complete reversal of the magnetization by $\mu_0 H_z\sim40$~mT. Inverting this procedure results in a symmetric reversal process (see supplementary Figures S3 and S4). 

Figure~\ref{fig2}j shows $B_{z}(x,y)$ averaged over two different areas of constant thickness, $B_z^\text{avg}$, as a function of $H_z$. These local hysteresis loops are shown for the 7 and 15-layer-thick regions indicated in Figure~\ref{fig2}a, though a full set can be found in supplementary Figure~S5. 
Regions of the sample thicker than about 9 layers show finite coercivity with a magnetic remanence of 70-80\% of the saturation magnetization $M_\text{sat}$, as exemplified by the data shown for the 15-layer-thick region.  In the thickest regions, with 13 layers or more, we also measure hysteresis loops with characteristic bow-tie-shapes, which is a sign of percolating magnetic domains, magnetic vortices, or skyrmion formation during magnetic reversal~\cite{soumyanarayanan_tunable_2017, avakyants_evidence_2023, brandao_observation_2019, ba_electric-field_2021}. 
In contrast, thinner regions of the sample with 7 layers or less, do not show measurable magnetic remanence and have magnetization curves characteristic of a soft ferromagnet.  
This behavior is consistent with previous observations in 6-layer flakes, although the measured saturation field of $\sim40$~mT is significantly smaller than the previously reported value of $\sim 0.6$~T~\cite{gong_discovery_2017}. 

\section{Layer-dependent magnetization texture}\label{sec3}
\begin{figure}[h]%
\centering
\includegraphics[width=0.9\textwidth]{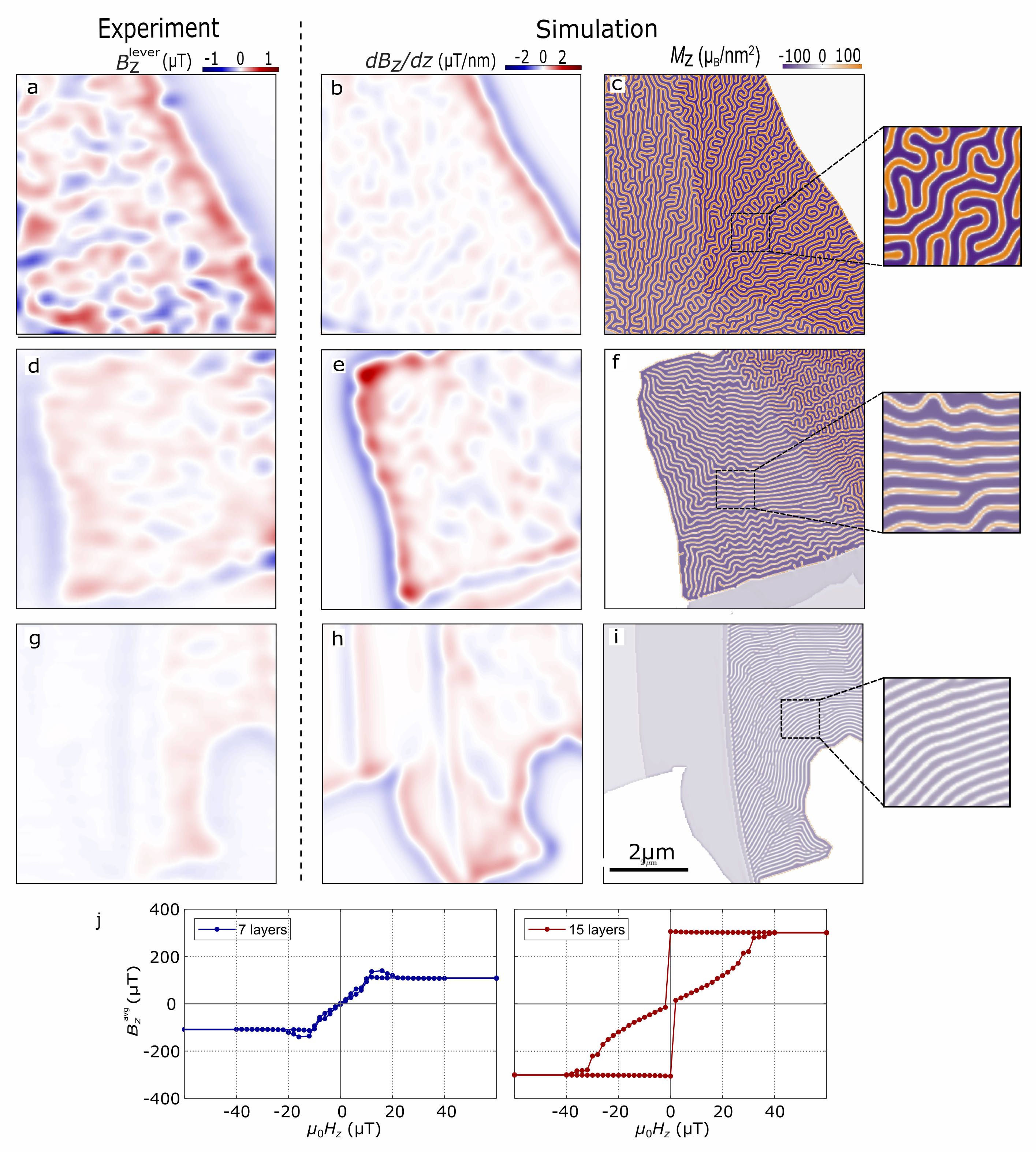}
\caption{Evolution of magnetic domains with thickness. (a) $B_z^\text{lever}(x, y)$ measured at $\mu_0H_z = 7.5$~mT over the 13 to 15-layer-thick region of the sample indicated by green and purple dots in Figure~\ref{fig1}a, together with (b) the corresponding simulation of $\frac{dB_z(x, y)}{dz}$ and (c) $M_z(x,y)$. (d-f) A similar measurement at the same $H_z$ over the 7 to 9-layer-thick region of the sample indicated by the aquamarine dot in Figure~\ref{fig1}a, together with corresponding simulations. (g-i) A final measurement at the same $H_z$ over the 3 to 5-layer-thick region indicated by the red and orange dots in Figure~\ref{fig1}a, together with corresponding simulations. (j) Simulated local hysteresis for 7 and 15 layers}\label{fig4}
\end{figure}
Our local magnetic measurements confirm that magnetic reversal in few-layer CGT depends on thickness. 
In order to shed light on the magnetization configurations corresponding to measured stray field patterns, we turn to micromagnetic simulations. We use a model based on the Landau-Liftshitz-Gilbert formalism, which is based on the geometry of the flake and known material parameters (see \textit{Methods} for details). We focus on three areas of the flake representing the behavior observed in the thick (13-15 layers), intermediate (7-9 layers), and thin regions (3-5 layers) of the sample. In our model, we use one less layer compared to the actual number of physical layers in the flake, based on our measurement of approximately one magnetically inactive layer.  In Figure~\ref{fig4}, we show $B_z^\text{lever}(x, y)$ measured above the sample with applied out-of-plane field $\mu_0H_z=7.5$~mT along with corresponding simulations of $\frac{dB_z(x, y)}{dz}$ and $M_z(x, y)$. As discussed in the \textit{Methods}, $B_z^\text{lever}(x,y) \propto \frac{dB_z(x, y)}{dz}$ and is measured by oscillating the cantilever on resonance and demodulating the SQUID-on-lever response at this frequency. It provides a more sensitive measure of small spatial features than $B_z(x,y)$.

Measured $B_z^\text{lever}(x, y)$ maps of all three regions agree well with the simulated $\frac{dB_z(x, y)}{dz}$, which reproduce the shapes, relative amplitudes, and characteristic lengths of the observed patterns. Furthermore, simulated magnetic hysteresis curves, shown in Figure~\ref{fig4}j, for the 7 and 15-layer-thick regions reproduce local hysteresis measurements shown in Figure~\ref{fig2}j. Given this agreement, we can turn to the corresponding simulated $M_z(x, y)$ configurations as the likely magnetization textures present in those regions of the sample. 

In particular, simulations of the thick region (13-15 layers), which is indicated by the green and purple dots in Figure~\ref{fig1}a, point to the presence of labyrinth domains, as shown in Figure~\ref{fig4}c. These structures are responsible for the bow-tie hysteresis behavior, which we observe in CGT flakes thicker than 13 layers, shown in Figure~\ref{fig2}j. 
A closer look at the simulations reveals that the labyrinth domains are separated by Neel-type domain walls in both surface layers, which gradually transform towards Bloch-type domain walls in the interior layers (supplementary Figure S11). 
Simulated $M_z (x,y)$ for the intermediate region with 7 to 9 layers, as indicated by the aquamarine dot in Figure~\ref{fig1}a, reveals long stripe domains (Figure~\ref{fig4}f). 
Simulated $M_z (x, y)$ for the thin region with 3 to 5 layers also shows stripe domains for 5 layers, but no domains for 3 layers. 
The absence of domains, in combination with soft ferromagnetic behavior in the very thin limit, is consistent with a previous report by Gong et al.\ \cite{gong_discovery_2017}. 

In our simulations, the presence of labyrinth domains in thick CGT as well as their transition to stripes, and -- eventually -- to regions without domains as thickness is reduced is robust to small adjustments of the material parameters and sample geometry. In fact, the strong layer dependence of the magnetization configuration is a direct result of the similar magnitudes of magnetocrystalline and shape anisotropy in the few-layer samples. As the thickness of the sample changes, the balance between magnetocrystalline anisotropy, which favors out-of-plane magnetization alignment, and shape anisotropy, which favors in-plane alignment, is altered.
The fact that this model reproduces the measured layer dependence suggests that the same mechanisms are at work in the experiment. 

However, in order to optimally match both simulated hysteresis curves and magnetic images with the corresponding measurements, both the magnetocrystalline anisotropy $K_\text{u}$ and the in-plane exchange stiffness $A_\text{ex}$ must be reduced when simulating the thinner parts of the flake compared to the thicker parts (see \textit{Methods}). The need to tune these parameters to achieve the best agreement suggests a dependence of $A_\text{ex}$ and $K_\text{u}$ on thickness. Although the mechanisms for such an effect remain unclear, a dependence of exchange interactions, and consequently $A_\text{ex}$, on the layer number has been predicted theoretically~\cite{fang_large_2018}. Another indication for such dependence is provided by the observation of a rapid drop in $T_\text{c}$ as a function of decreasing thickness in CGT~\cite{gong_discovery_2017}.

\section{Skyrmionic magnetization texture}\label{sec3}
\begin{figure}[h]%
\centering
\includegraphics[width=1\textwidth]{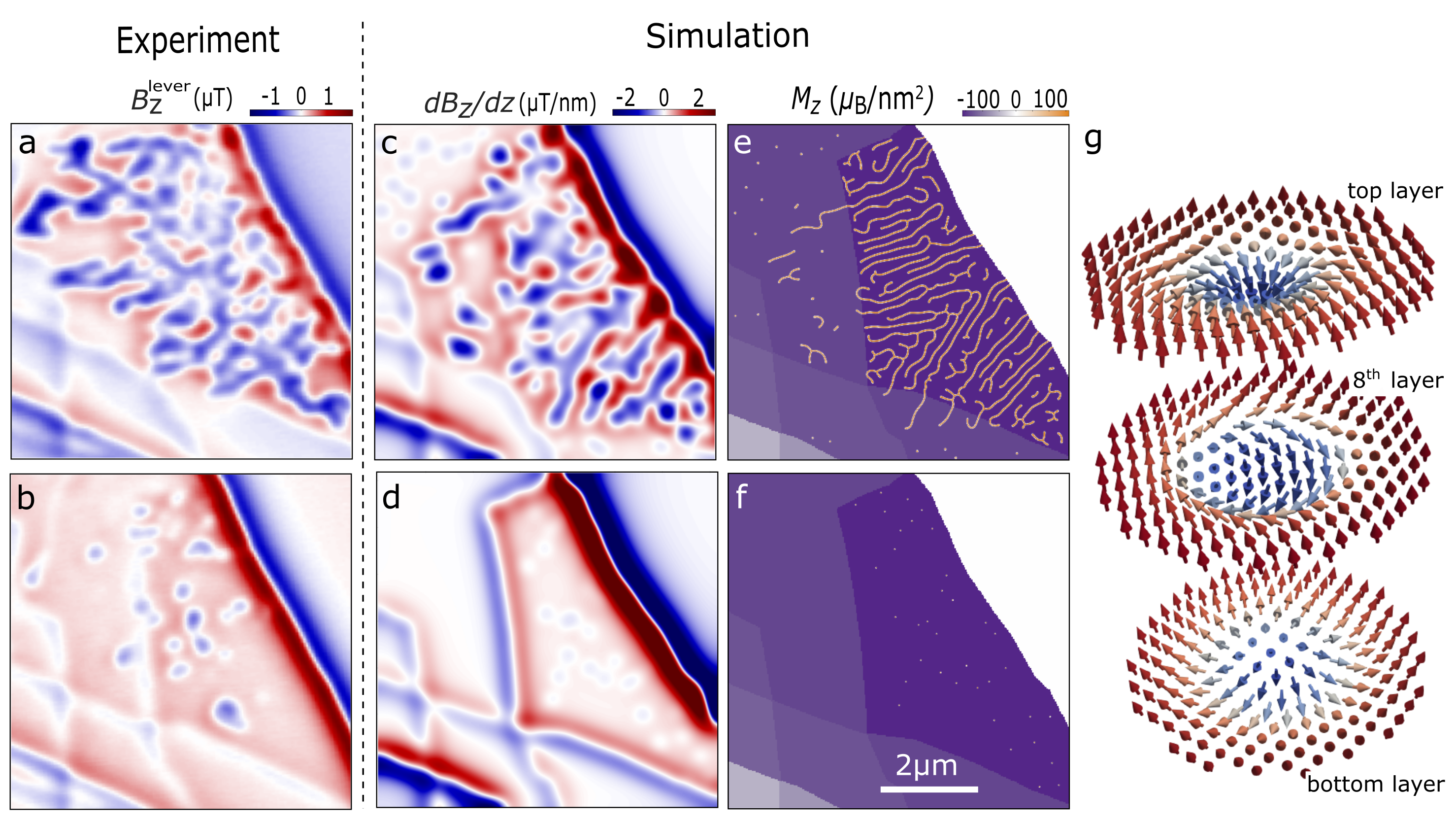}
\caption{Skyrmionic magnetization texture. (a) and (b) $B_z^\text{lever} (x,y)$ maps of the thick part of the field-cooled sample at $\mu_0H_z = 17$~mT and $28$~mT, respectively.  (c-f) Corresponding $\frac{dB_z(x, y)}{dz}$ and magnetization $M_z (x,y)$ simulated at $\mu_0H_z = 26$~mT and $38$~mT, respectively. (g) The simulated magnetization configuration of the magnetic bubbles at $\mu_0H_z = 28$~mT, showing a Bloch-type skyrmionic texture in the middle layer, which gradually transforms into a N\'eel-type texture at the surface layers}\label{fig5}
\end{figure}
Simulations of the thick part of the sample (12 to 15 layers thick) reveal labyrinth domains with similar magnetization patterns as were previously observed in flakes in the range of 100 nm or thicker~\cite{han_topological_2019}. In those experiments, it was observed that under increasing $H_z$, the labyrinth domains transform into bubbles that are mostly homochiral.
Such bubbles are topologically identical to skyrmions and are characterized by a topological charge of ±1. 
Topologically non-trivial spin textures like skyrmions are usually stabilized in non-centrosymmetric magnetic systems via a Dzyaloshinskii-Moriya interaction (DMI)~\cite{bogdanov_physical_2020}, which is not present in this system. 
Nevertheless, the competition between uniaxial anisotropy and magnetostatic energy can give rise to a plethora of magnetic patterns including labyrinth domains, stripes, and skyrmionic bubbles.~\cite{hubert_magnetic_1998, nielsen_magnetic_1979}. 

To determine whether such skyrmionic bubbles are found in few-layer-thick CGT, we field-cool the sample at $\mu_0H_z = 10\,\text{mT}$.  
Figure~\ref{fig5}a shows a measurement of $B_z^\text{lever} (x,y)$ over the thick part of the field-cooled sample after the applied field was increased to $\mu_0H_z =17$~mT. Stray field patterns characteristic of labyrinth domains are visible over most of the 15-layer-thick region with some percolating features expanding to neighboring regions. As shown in Figure~\ref{fig5}b, after the field is further increased to $\mu_0H_z = \SI{28}{\milli\tesla}$, the patterns related to labyrinth domains shrink and transform into bubble-like features. 

In corresponding micromagnetic simulations, we mimic the field-cooling procedure by starting with an arbitrary labyrinth domain, applying $\mu_0H_z = \SI{10}{\milli\tesla}$, and then letting the system relax to find the minimum energy state. The resulting simulated $\frac{dB_z(x, y)}{dz}$ and $M_z (x,y)$ in Figure~\ref{fig5}c and~\ref{fig5}e present features similar to those measured in the field-cooled $B_z^\text{lever}(x,y)$. The corresponding $M_z(x, y)$ shows that stripe domains start from the edge of the flake, form multiple branches, and expand to the thinner regions. 
As the field is increased in the simulations, the corresponding $\frac{dB_z(x, y)}{dz}$ map in Figure~\ref{fig5}d qualitatively matches the bubble-like features in the measured $B_z^\text{lever}(x,y)$. $M_z (x, y)$ maps reveal that, upon the increase of $H_z$, the underlying domains shrink and transform into bubbles, most of which have skyrmionic magnetization texture, as shown in Figure~\ref{fig5}f. As was observed in our magnetization simulations of the labyrinth domains, whose helicity transforms through the thickness of the flake, the skyrmionic bubbles also show a modulating helicity from the top to the bottom layer, as schematically shown in Figure~\ref{fig5}g. This behavior has been previously proposed in systems with skyrmions \cite{zhang_direct_2018} and has also been observed in skyrmionic bubbles \cite{yao_chirality_2022}. Such bubbles can also be obtained by zero-field-cooling this sample (see supplementary Figure S4).

%
\section{Conclusion}\label{sec13}
We have investigated the thickness dependence of magnetic ordering in CGT down to the few-layer limit. The measured magnetic hysteresis for different numbers of layers reveals a transition from a bow-tie hysteretic behavior to a soft behavior without remanence for less than 8 layers. Comparison with micromagnetic simulations indicates that complex stray field patterns observed for CGT thicker than 5 layers result from labyrinth and stripe-like magnetization configurations and -- under some conditions -- from skyrmionic bubbles. These complex magnetic textures emerge from the competition between magnetocrystalline anisotropy and magnetostatic interactions. Our experimental results are reproduced by micromagnetic simulations under the assumption that the magnitude of these two energies is similar. The agreement is further optimized in the thinnest regions of the sample  assuming a decreasing magnetocrystalline anisotropy and exchange stiffness with the decreasing number of layers. Although the mechanism for this dependence of the material's magnetic properties is unclear, the decrease in exchange stiffness with decreasing number of layers has been suggested theoretically~\cite{fang_large_2018}. Nevertheless, we cannot exclude that external factors, which could also vary as a function of sample thickness, including level of oxidation or defect density, could be responsible.

\section{Methods}\label{sec11} 
\subsection{$\text{Cr}_2\text{Ge}_2\text{Te}_6$/hBN heterostructure fabrication}

The heterostructure is fabricated in ambient conditions using a dry viscoelastic stamping method~\cite{castellanos-gomez_deterministic_2014}. 
CGT flakes with different thicknesses are first mechanically exfoliated from a single crystal (HQ graphene) onto a SiO$_2$/Si substrate with pre-patterned number markers. 
A 10-nm-thick hBN flake is exfoliated from the bulk form on the Polydimethylsiloxane (PDMS) thin film and then stacked on top of the CGT flake for passivation. 
To minimize degradation of the CGT surface, the stacking process is carried out below 120$^\circ$C in all steps and handled within less than 10~minutes~\cite{lohmann_probing_2019}. 

\subsection{SQUID-on-lever fabrication}

We pattern a nanometer-scale SQUID via focused-ion-beam (FIB) milling at the apex of a cantilever coated with Nb \cite{wyss_magnetic_2022}.
The top side of the cantilever is deposited with 50 nm of Nb and 20 nm of Au to create a superconducting film with enhanced resistance against ion implantation effects of subsequent Ga\(^+\)-FIB milling steps.
Before film deposition, FIB milling is used to create a 650-nm-wide plateau on the cantilever's tip. 
After film deposition, FIB milling is used to separate two superconducting leads across the cantilever and to create a superconducting loop with a diameter of 80~nm with two constriction-type Josephson junctions on the plateau. 
The SQUID is characterized and operated at 4.2~K in a semi-voltage biased circuit and its current \(I_\text{SQ}\) is measured by a series SQUID array amplifier (Magnicon).  The effective diameter of the SQUID extracted from the interference pattern (supplementary Figure S2) is 270~nm.
The SQUID attains a DC magnetic flux sensitivity of  \(S_\phi = 1 \,\mu\Phi_0/\sqrt{\text{Hz}}\) and an AC flux sensitivity at 10~kHz of \(S_\phi = 0.2 \,\mu\Phi_0/\sqrt{\text{Hz}}\) and remains sensitive up to \(\mu_0 H_z > 250 \,\text{mT}\) (see supplementary S2). 

\subsection{Hybrid imaging}

The SOL scanning probe is capable of simultaneously performing AFM and SSM.
It operates in a custom-built scanning setup under high vacuum in a \(^4\text{He}\) cryostat.
Non-contact AFM is carried out using a fiber-optic interferometer to measure the cantilever displacement and a piezo-electric actuator driven by a phase-locked loop (PLL) to resonantly excite the cantilever at \(f_0 = 285.28 \, \text{kHz}\) to an amplitude \(\Delta z = 16 \, \text{nm}\). 
%
%
Since the current response of the SQUID-on-lever is proportional to the magnetic flux threading through it, this response provides a measure of the z-component of the local magnetic field integrated over the loop. By scanning the sample using piezoelectric actuators at a constant tip-sample spacing of 200~nm, we map $B_z (x, y)$. 
We can also measure \(B_\text{lever} \propto \text{d}B_z/\text{d}z \) by demodulating the SQUID-on-lever response at the cantilever oscillation frequency. Due to spectral filtering, the resulting signal contains less noise than DC measurements of $B_z$. 
The spatial resolution of the SSM is limited by the tip-sample spacing and by the 270-nm SQUID-on-lever effective diameter. 
Maps of $B_z(x, y)$ and $B_\text{lever}(x,y)$ are taken using a scanning probe microscopy controller (Specs) at a scan rate of 330~$\text{nm}/\text{s}$, 338~ms per pixel, and typically take several hours.

%
%
 
\subsection{Micromagnetic simulations}

We simulate the flake's magnetization configuration using the \textit{Mumax$^3$} software package~\cite{vansteenkiste_design_2014, exl_labontes_2014}. The software utilizes the Landau-Lifshitz-Gilbert micromagnetic formalism with finite-difference discretization.
To mimic the layered structure of CGT we make use of the finite difference mesh by setting the thickness of a mesh cell to the thickness of a CGT layer. The geometry, estimated from optical images, and corresponding number of layers, is reproduced for three different parts of the studied CGT flake. The structure is discretized into cells of size $3.5 \text{ nm}\times 3.5 \text{ nm}\times 0.7 \text{ nm}$.
The saturation magnetization is chosen to be $M_\text{sat} = 2.2 \, \mu_\text{B}/\text{Cr}$, based on the linear fit of the magnetization presented in Fig~\ref{fig1}f.  In order to reproduce the observed presence of domains in the thicker and their absence in the thinner parts of the flake, we assume that the magnetocrystalline anisotropy $K_\text{u}$ of the system is similar to its shape anisotropy $K_\text{s}$~\cite{yu_magnetic_2012, belliard_stripe_1997} which can be approximated as an infinitely extended plate: $K_s = 1/2 \, \mu_0 M_\text{sat}^2 \approx 13600 \,\text{J}/\text{m}^3$ \cite{hubert_magnetic_1998}.
The intra-layer exchange stiffness $A_\text{ex}$ is estimated based on the Curie temperature $T_\text{c}$~\cite{coey_magnetism_2010, gong_discovery_2017}.
Both $K_\text{u}$ and $A_\text{ex}$ were further optimized from this starting point to optimally match our both measured hysteresis curves and magnetic images. Two values were used for the $K_\text{u} = 13700 \, \text{J}/\text{m}^3$ for the simulation of the thicker part of the sample (11-15 layers) and $K_\text{u} = 12800 \,\text{J}/\text{m}^3$ for the thinner parts (3-9 layers). This change was implemented since domains start forming at different $H_z$ for the thinner and thicker parts. The intra-layer exchange stiffness  was also modulated for the three different simulated regions with $A_\text{ex} =2.5\cdot 10^{-13}$~J/m, $1.8\cdot 10^{-13}$~J/m and $1.4\cdot 10^{-13}$~J/m for the thicker  (12-15 layers), intermediate (7-9 layers) and thinner (3-5 layers) regions. This was again done to match measured magnetic hysteresis and respective magnetic field maps. 
The interlayer exchange stiffness is assumed to be $3\,\%$ of the intralayer stiffness, estimated based on the ratio of the interlayer and intralayer exchange interactions between Cr atoms~\cite{zhu_topological_2021}. Small adjustments on the interlayer exchange stiffness do not change the magnetic behavior of the simulated flakes. 

In order to generate simulated maps of $B_z(x,y)$ and $\frac{dB_z(x, y)}{dz}$, we use the magnetization maps generated by \textit{Mumax$^3$}. We calculate $B_z(x,y)$ at a height of 200~nm above the sample, corresponding to our SQUID-sample distance. $\frac{dB_z(x, y)}{dz}$ is calculated assuming $dz=\SI{16}{\nano\meter}$ corresponding to the oscillation amplitude of the cantilever. Finally, we apply a Gaussian blurring of $2\sigma=\SI{270}{\nano\meter}$ to approximate the point-spread function of the SQUID sensor.

%
\subsection{Author Contributions}

M.P.,  K.B., and A.V. conceived the project. M.D. and T.K.C. fabricated the sample. K.B., A.V., and D.J. performed the experiment. K.B., D.J., and A.V. analyzed the data. A.V. and B.G. performed the micromagnetic simulations. M.P. and K. B.  wrote the manuscript with input from A.V. and D.J.  All authors discussed the results and commented on the manuscript.

\bmhead{Acknowledgments}
We thank Dr. David Broadway and Prof. Patrick Maletinsky for fruitful discussions and assisting with the magnetization reconstruction. We also thank Sascha Martin and his team in the machine shop of the Department of Physics at the University of Basel for their role in building the scanning probe microscope. Calculations were performed at sciCORE (http://scicore.unibas.ch/) scientific computing center at the University of Basel. We acknowledge support of the European Commission under H2020 FET Open grant “FIBsuperProbes” (Grant No. 892427), the SNF under Grant No. 200020-207933, the Canton Aargau, and the Novo Nordic Foundation grant "Superior" (Grant No. NNF21OC0068015).


\end{document}


\title[Supplementary: Vervelaki et al.]{Supplementary Information:\\ Visualizing thickness-dependent magnetic textures in few-layer $\text{Cr}_2\text{Ge}_2\text{Te}_6$}


\author{A.~Vervelaki} \affiliation{Department of Physics, University of Basel, 4056 Basel, Switzerland}

\author{K.~Bagani} \affiliation{Department of Physics, University of Basel, 4056 Basel, Switzerland}

\author{D.~Jetter} \affiliation{Department of Physics, University of Basel, 4056 Basel, Switzerland}

\author{M.-H.~Doan} \affiliation{Department of Physics, Technical University of Denmark, 2800 Kongens Lyngby, Denmark}

\author{T.~K.~Chau} \affiliation{Department of Physics, Technical University of Denmark, 2800 Kongens Lyngby, Denmark}

\author{B.~Gross} \affiliation{Department of Physics, University of Basel, 4056 Basel, Switzerland}

\author{D.~Christensen} \affiliation{Department of Energy Conversion and Storage, Technical University of Denmark, 2800 Kongens Lyngby, Denmark}

\author{P.~Bøggild} \affiliation{Department of Physics, Technical University of Denmark, 2800 Kongens Lyngby, Denmark}

\author{M.~Poggio} \affiliation{Department of Physics, University of Basel, 4056 Basel, Switzerland} \affiliation{Swiss Nanoscience Institute, University of Basel, 4056 Basel, Switzerland}

\maketitle
\newpage

\section{Atomic force microscopy} \label{secA1}

The SOL scanning probe is capable of performing conventional non-contact mode AFM.
This allows for a topographic mapping of the measured CGT flake in parallel to SSM.
Fig. S1a shows a greyscale optical image of the investigated CGT flake with lines indicating corresponding AFM line-cuts plotted below.
Fig. S1b shows the topography of the flake gathered via SOL AFM by dynamically adjusting the tip-sample spacing to keep a constant frequency shift of -23~Hz correlated to 150~nm tip-sample distance. 
The AFM image shows the regions of different thickness across the flake.
A surface roughness of 300~pm allows us to measure single atomic layer transitions of CGT with \(\Delta z = 0.7\,\text{nm}\).
Fig. S1c shows corresponding line cuts extracted from the AFM image showing the respective layer transitions across the flake.
%
\vspace{20mm}

\begin{figure}[H]%
\renewcommand{\thefigure}{S1}
\centering
\includegraphics[width=0.7\textwidth]{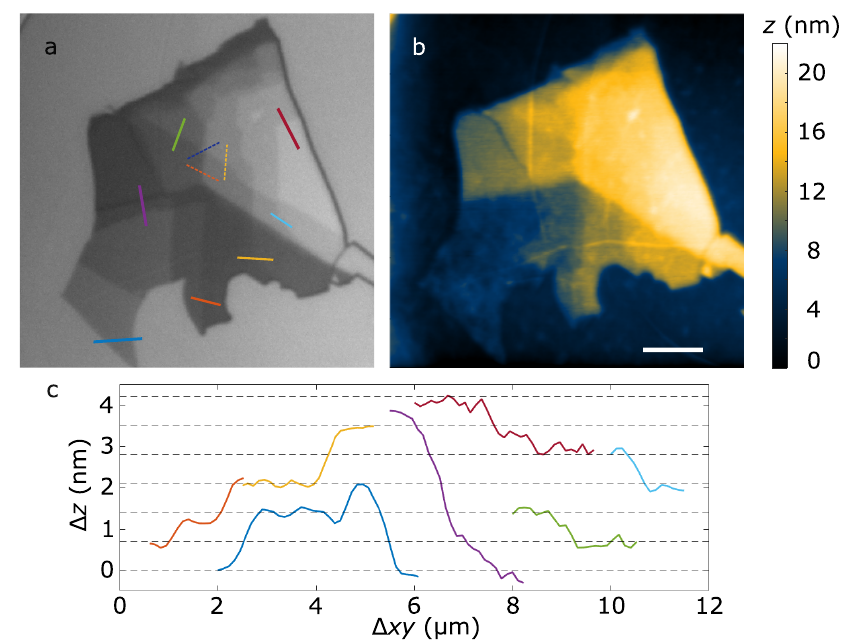}
\caption{Greyscale optical image of the investigated CGT flake (a). Solid lines indicate measured layer transitions shown in (c). (b) Topography of the CGT flake measured with the SOL, 
and (c) corresponding AFM line scan layer transitions with dashed lines indicating 0.7~nm CGT layer transitions.}\label{figA1_1}
\end{figure}

\newpage

\section{SQUID characteristics}\label{secA2}

\begin{figure}[H]%
\renewcommand{\thefigure}{S2}
\centering
\includegraphics[width=1\textwidth]{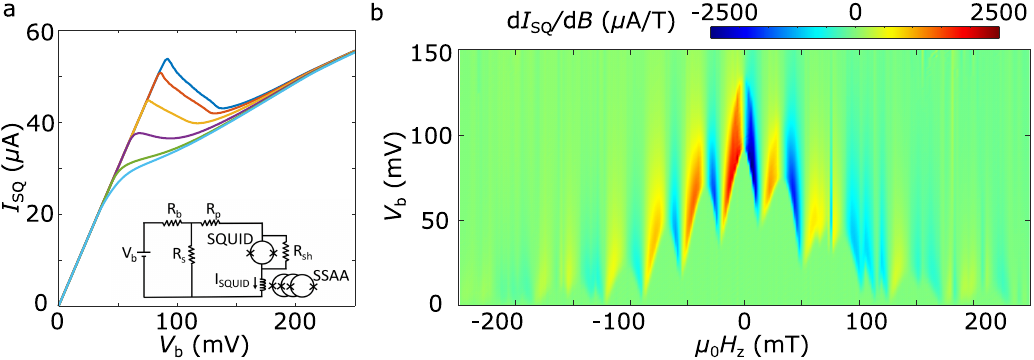}
\caption{(a) Circuit scheme and current-voltage curves at applied out-of-plane magnetic fields \(\mu_0 H_z\) of -2~mT, -6~mT, -10~mT, -14~mT, -18~mT and -22~mT. (b) Map or the SQUID's magnetic response d\(I_\text{SQ}/\)d\(B\) in a field range of -250~mT to 250~mT.}\label{figA2_1}
\end{figure}

The SQUID is characterized and operates in a semi-voltage biased circuit.
Its current \(I_\text{SQ}\) is measured by a Magnicon series SQUID array amplifier as shown in the inset of Fig.~S2a.
Sweeping the bias voltage \(V_\text{b}\) at various out-of-plane applied magnetic fields \(\mu_0 H_z\) creates I-V curves as shown in Fig.~S2a and a full set gives a map of the magnetic response \(dI_\text{SQ} / dB\) as shown in Fig.~\ref{figA2_1}b.
The observed modulation yields the SQUID's effective area with an approximate diameter of 270~nm.
The SQUID response reaches up to d\(I_\text{SQ}/\)d\(B\) = 3000 \(\mu\)A/T and persists beyond \(B_\text{a,z}>250\)~mT.
In spatial scans, the SQUID DC signal resolves magnetic signatures down to $1\,\mu\text{T}$.
\vspace{20mm}

%

\newpage
\section{Magnetic Imaging}\label{secA3}
\begin{figure}[H]%
\renewcommand{\thefigure}{S3}
\centering
\includegraphics[width=1\textwidth]{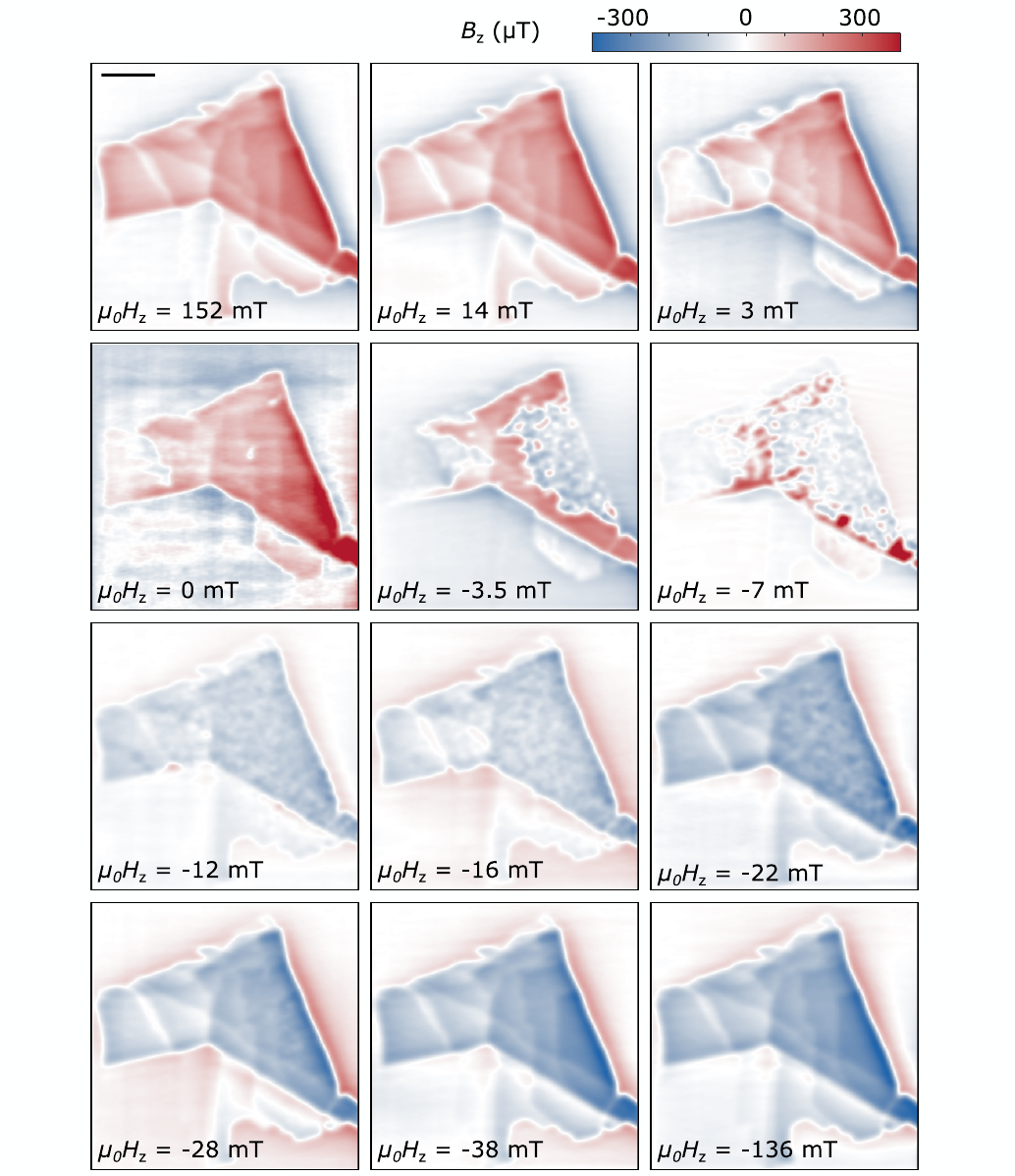}
\caption{Selection of images of the magnetic field \(B_z(x,y)\)  with progressively decreasing \(H_z\). 
The scale bar in the top-left corresponds to $5\,\mu\text{m}$.}\label{figA4_outofplaneBdc}
\end{figure}

\begin{figure}[H]%
\renewcommand{\thefigure}{S4}
\centering
\includegraphics[width=1\textwidth]{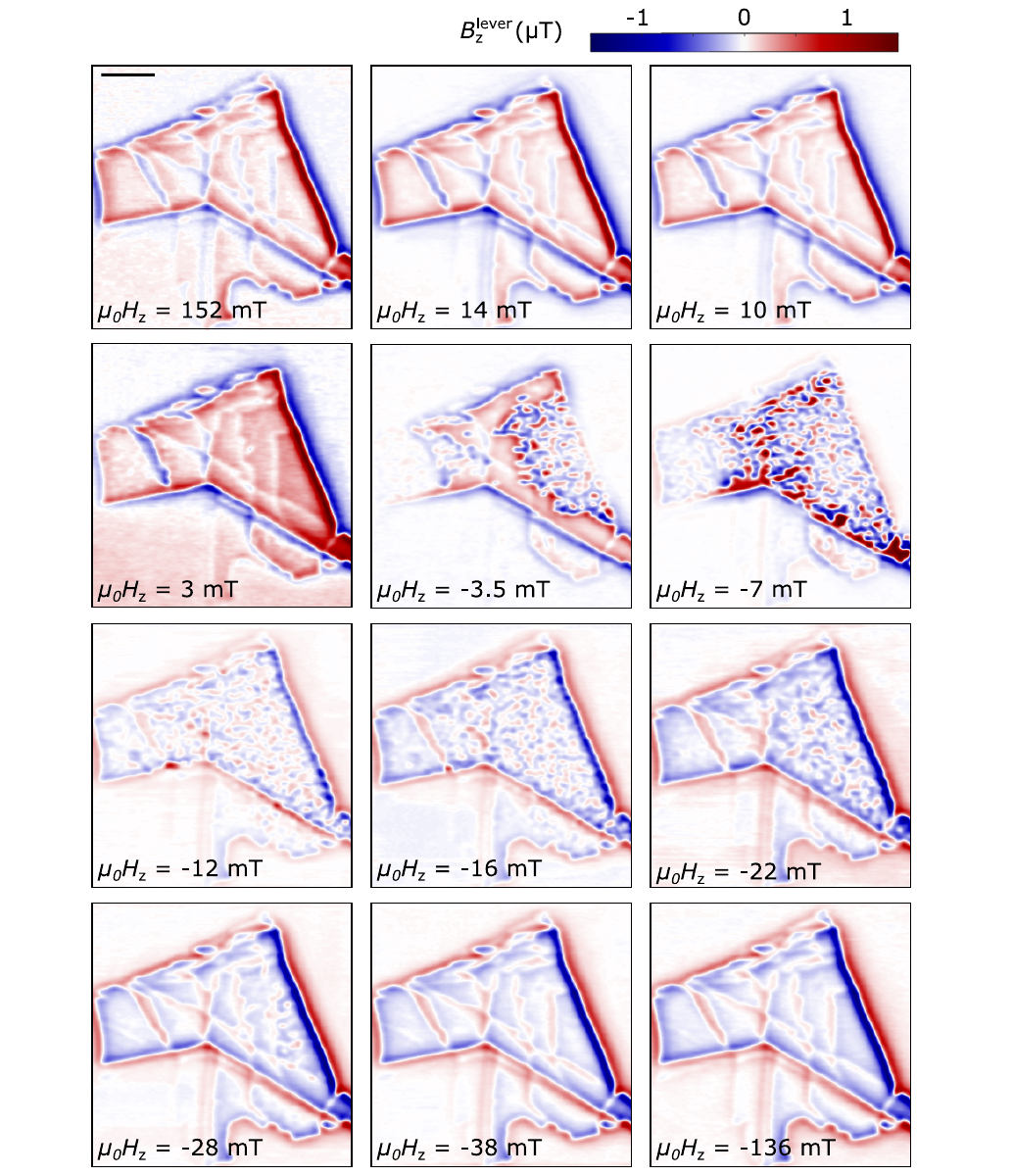}
\caption{Selection of images of the magnetic field gradient signal \(B_z^\text{lever}(x,y)\)  with progressively decreasing \(H_z\). 
The scale bar in the top-left corresponds to $5\,\mu\text{m}$.} \label{figA4_outofplaneBac}
\end{figure}

\begin{figure}[H]%
\renewcommand{\thefigure}{S5}
\centering
\includegraphics[width=.9\textwidth]{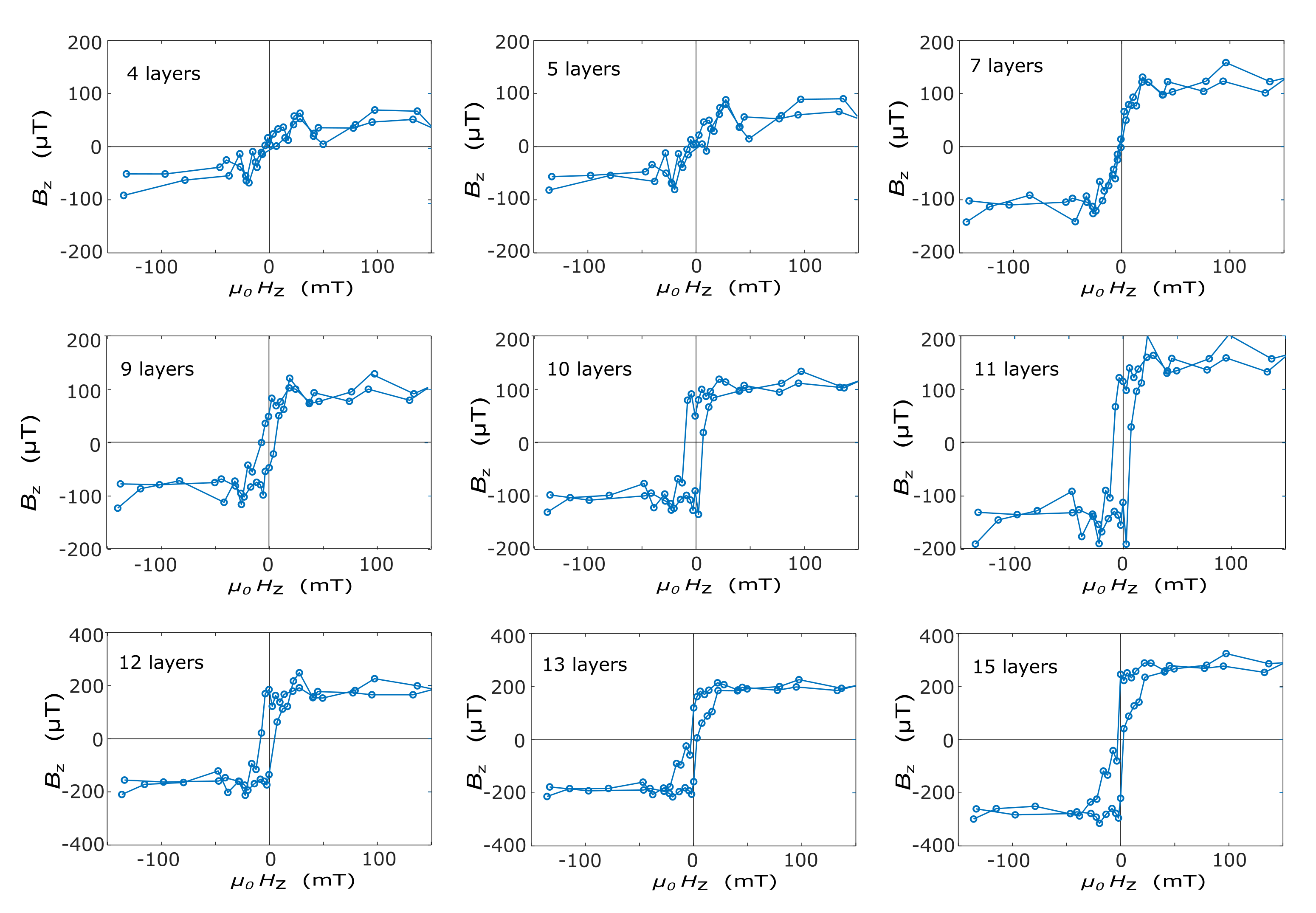}
\caption{Selection of \(B_z\) averaged over a region of constant thickness, indicated on the plot, plotted as a function of out-of-plane applied field \(H_z\).}\label{figA4_hist}
\end{figure}

We image a pristine CGT flake by measuring the DC out-of-plane magnetic field \(B_z\) and \(B_\text{lever} \propto \text{d}B_z/\text{d}z\).
We imaged the flake over two full hysteresis loops of out-of-plane and in-plane applied magnetic field, in order to distinguish layer-dependent switching behavior and domain formation.
Figures \ref{figA4_outofplaneBdc} and \ref{figA4_outofplaneBac} show a selection of \(B_z\) and \(B_\text{lever}\) images of a downward sweep of the applied out-of-plane field \(H_z\).
Figure \ref{figA4_hist} shows local hysteresis curves for regions of different thickness throughout the CGT flake. In order to obtain curves related to local magnetic behavior, $B_z(x,y)$ is averaged over a region of constant thickness and then plotted as a function of \(H_z\).

\begin{figure}[H]%
\renewcommand{\thefigure}{S6}
\centering
\includegraphics[width=1\textwidth]{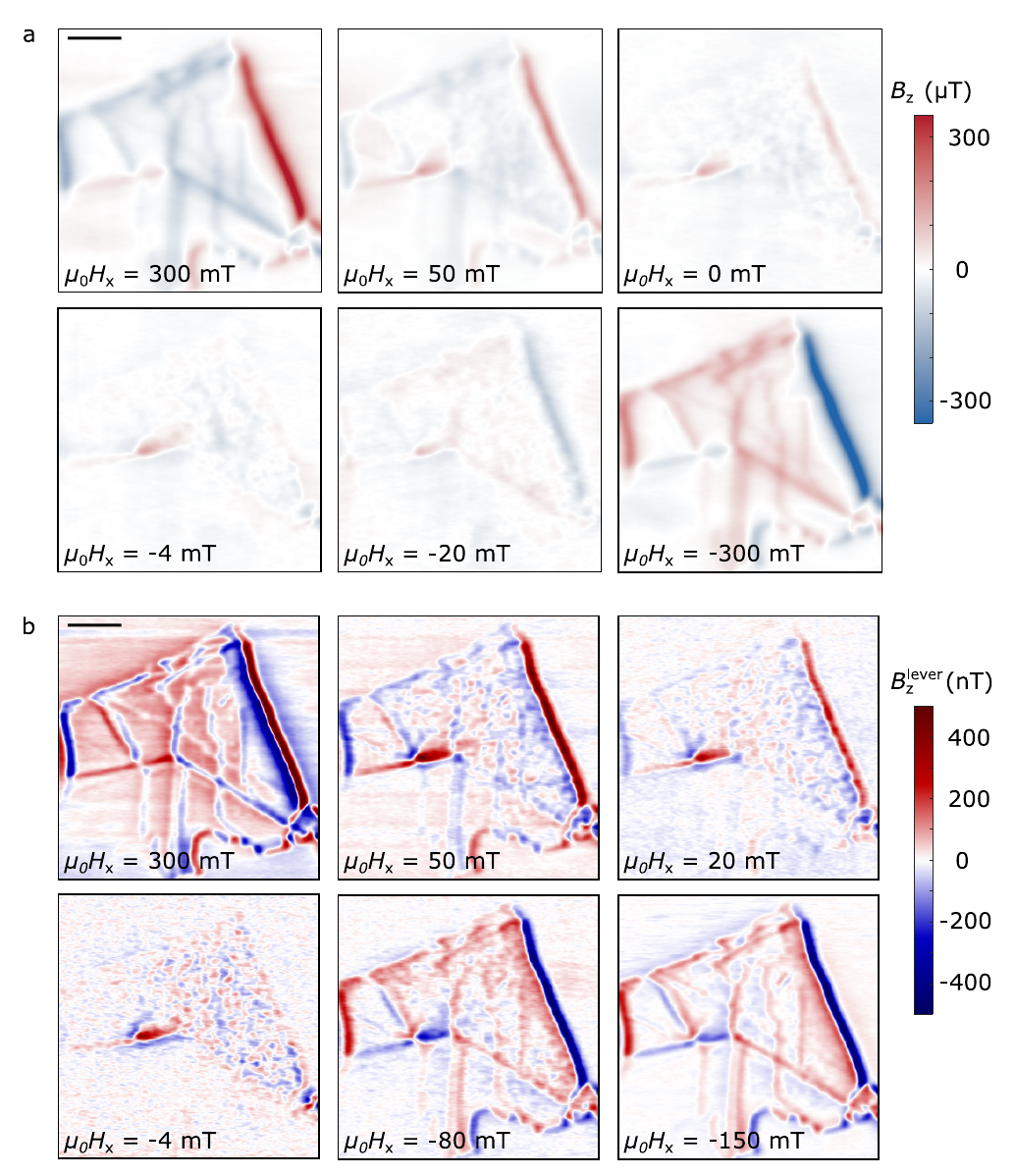}
\caption{Selection of images of the magnetic field \(B_z(x,y)\) (a) and the magnetic field gradient signal \(B_z^\text{lever} (x, y)\) (b) with progressively decreasing in-plane applied magnetic field \(H_x\) and constant out-of-plane field \(\mu_0 H_z = -2\)~mT.
Scale bars in the top-left correspond to $5\,\mu\text{m}$.}\label{figA4_inplane}
\end{figure}

Figure \ref{figA4_inplane} shows a selection of images of a downward sweep of the applied in-plane field \(H_x\).
First, the sample is initialized in the saturated state at $\mu_0H_x=-400$~mT. As $H_x$ is stepped towards zero field, small domains form across the entire flake and then gradually polarize in the opposite direction as the field is reversed.
We observe no magnetic remanence and a saturation field of $H_k\simeq 150$~mT, being consistent with an out-of-plane magnetic anisotropy.

\newpage
\newpage
%
\begin{figure}[H]%
\renewcommand{\thefigure}{S7}
\centering
\includegraphics[width=1\textwidth]{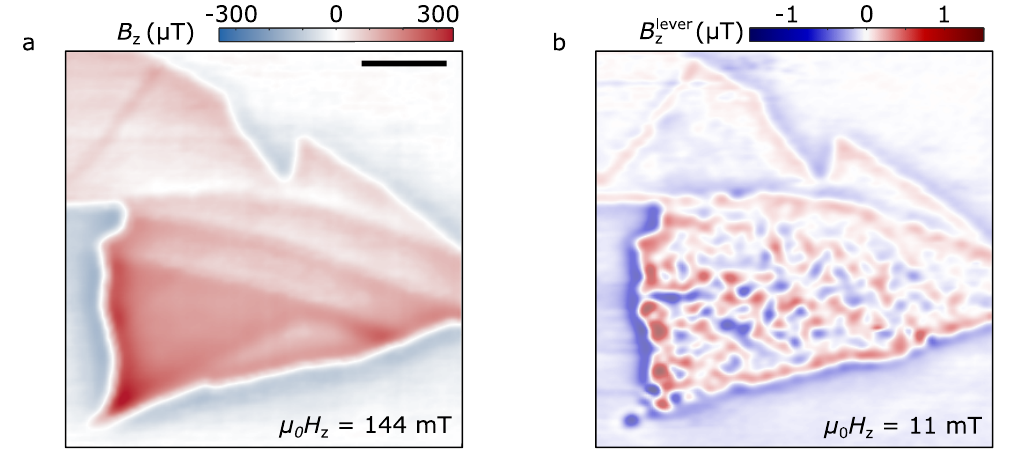}
\caption{Magnetic field maps of another CGT flake. (a) $B_z(x,y)$ of the flake saturated by an out-of-plane magnetic field. (b) \(B_z^\text{lever}(x,y)\) after reversing the applied field.
The scale bar corresponds to 3~$\mu$m.}\label{figA4_thick}
\end{figure}

\newpage
Our study also contains images of another part of a CGT flake with a thickness between 3 to 7 atomic layers.
Figure~\ref{figA4_thick}~(a) shows \(B_z(x,y)\) of the flake with in an applied out-of-plane magnetic field \(\mu_0 H_z = 144\)~mT.
Magnetization reconstructions combined with the number of layers similar to Figure~1f give the same saturation magnetization of \(M_\text{sat} = 2.2 \, \mu_\text{B}/\text{Cr}\) atom.
Figure~\ref{figA4_thick}~(b) shows \(B_z^\text{lever}(x,y)\) of the flake after switching the applied out-of-plane magnetic field from negative to positive field of \(\mu_0 H_z = 11\)~mT.
As in the flake presented in Figure~2, this flake's thicker parts show complex domain formations whereas its thinner parts show a soft ferromagnetic behavior.
\vspace{20mm}
%
%
\newpage
\section{Image processing}\label{secA4}
Fourier-space image processing is employed to filter noise, remove artifacts from the scanning process, and reconstruct the magnetization from measurements of the stray magnetic field.
We use a Hanning window to filter high-frequency noise and line scan artifacts. 
The sample magnetization \(M(x,y)\) is reconstructed from \(B_z(x,y,z)\) maps via a backpropagation in k-space \(M = \mathbb{A}^{-1} B_z\).
Given a thin and fully polarized magnetic flake, we assume a homogeneous magnetization direction confined to a 2-dimensional plane.
The flake generates a magnetic field with out-of-plane component \(B_z\) measured in a plane at spacing \(z\). 
A magnetization fully pointing out-of-plane is defined in k-space as \(\mathcal{M}_z = \alpha \mathcal{B}_z / k\)  
with \(\alpha = 2 e^{kz} / \mu_0\) considering the propagation of the field from its source to the measured plane~\cite{broadway_improved_2020}.
%
%
\newpage
\section{Micromagnetic simulations}\label{secA5}
\begin{figure}[H]%
\renewcommand{\thefigure}{S8}
\centering
\includegraphics[width=0.7\textwidth]{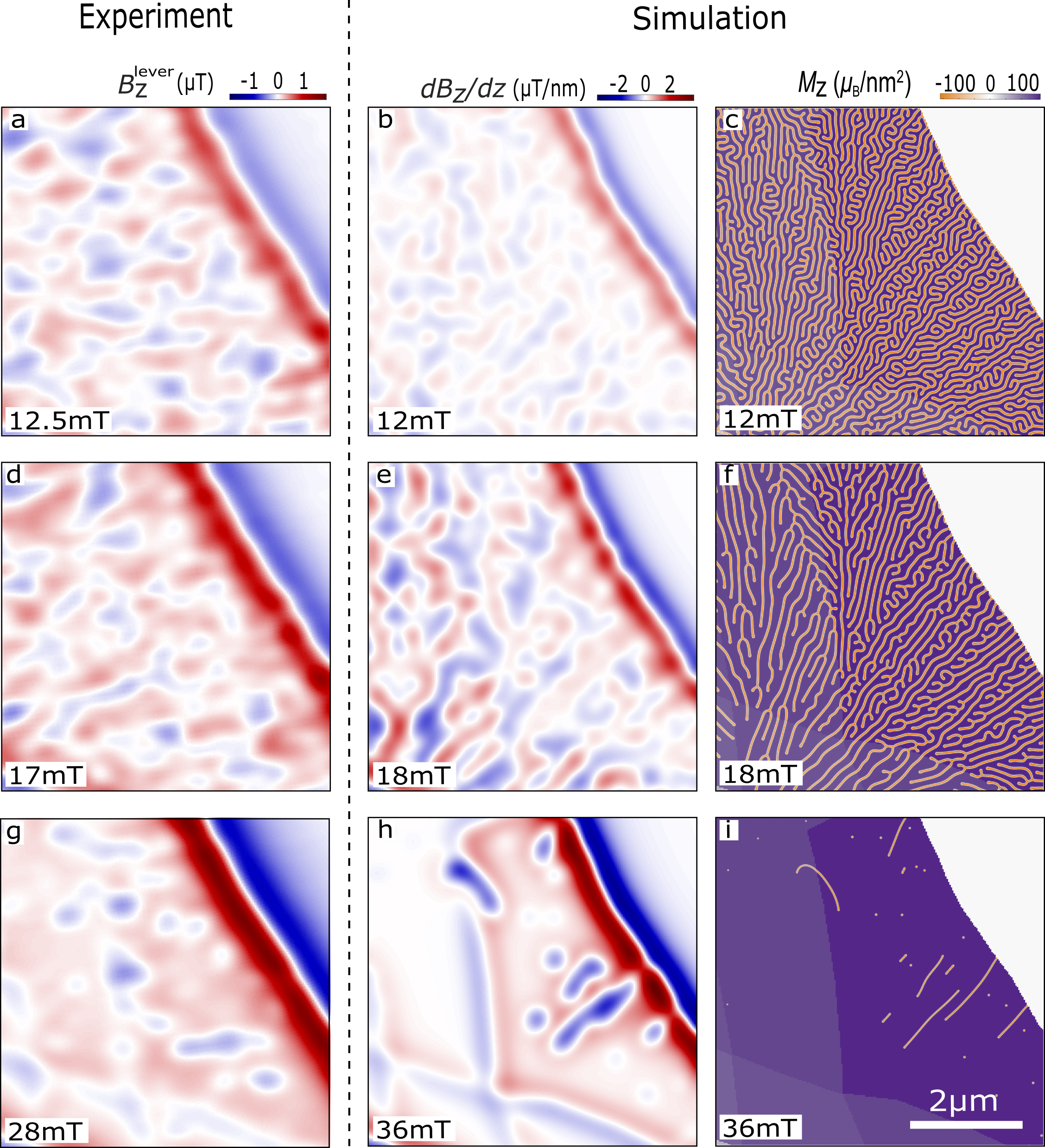}
\caption{Evolution of magnetic domains under increasing  $H_z$ for 13-15 layers. (a) $B_z^\text{lever}(x, y)$ measured at $\mu_0H_z = 12.5$~mT over the 13 to 15-layer-thick region of the sample indicated by green and purple dots in Figure 1a, together with (b) the corresponding simulation of $dB_z/dz(x, y)$ and (c) $M_z(x,y)$ under similar $\mu_0H_z$. (d-f) and (g-i) show again the measured $B_z^\text{lever}(x, y)$ together with corresponding simulations under externally applied $\mu_0H_z = 17$~mT and $\mu_0H_z = 28$~mT respectively.}\label{figS4_sim_15L}
\end{figure}

\begin{figure}[H]%
\renewcommand{\thefigure}{S9}
\centering
\includegraphics[width=0.7\textwidth]{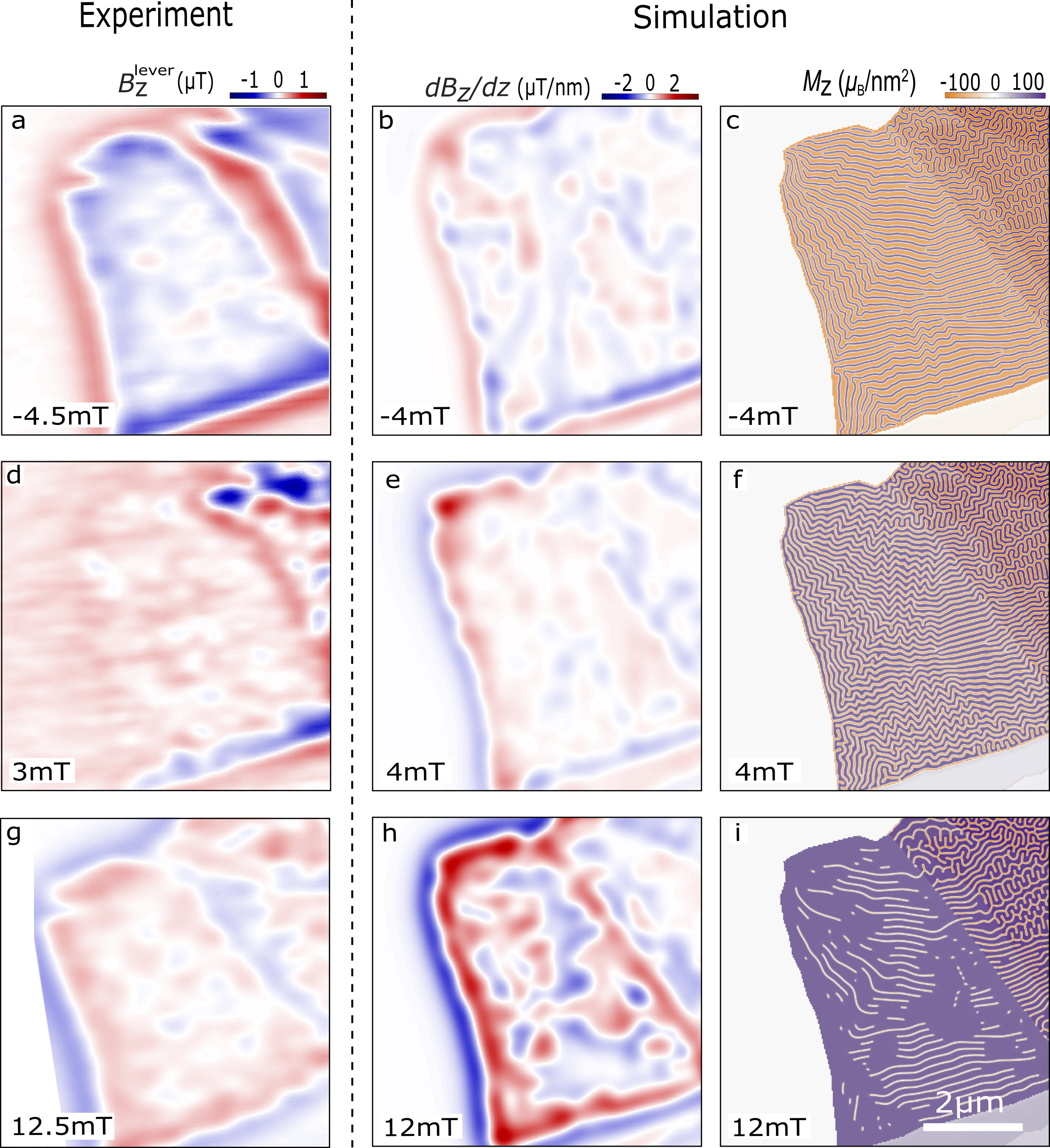}
\caption{Evolution of magnetic domains under increasing  $H_z$ for 7-9 layers.  (a) $B_z^\text{lever}(x, y)$ measured at $\mu_0H_z = -4.5$~mT over the 7 to 9-layer-thick region of the sample indicated by green and purple dots in Figure~1a, together with (b) the corresponding simulation of $dB_z/dz(x, y)$ and (c) $M_z(x,y)$ under similar $\mu_0H_z$. (d-f) and (g-i) show again the measured $B_z^\text{lever}(x, y)$ together with corresponding simulations under externally applied $\mu_0H_z = 3$~mT and $\mu_0H_z = 12.5$~mT respectively.}\label{figS4_sim_7L}
\end{figure}

\begin{figure}[H]%
\renewcommand{\thefigure}{S10}
\centering
\includegraphics[width=0.7\textwidth]{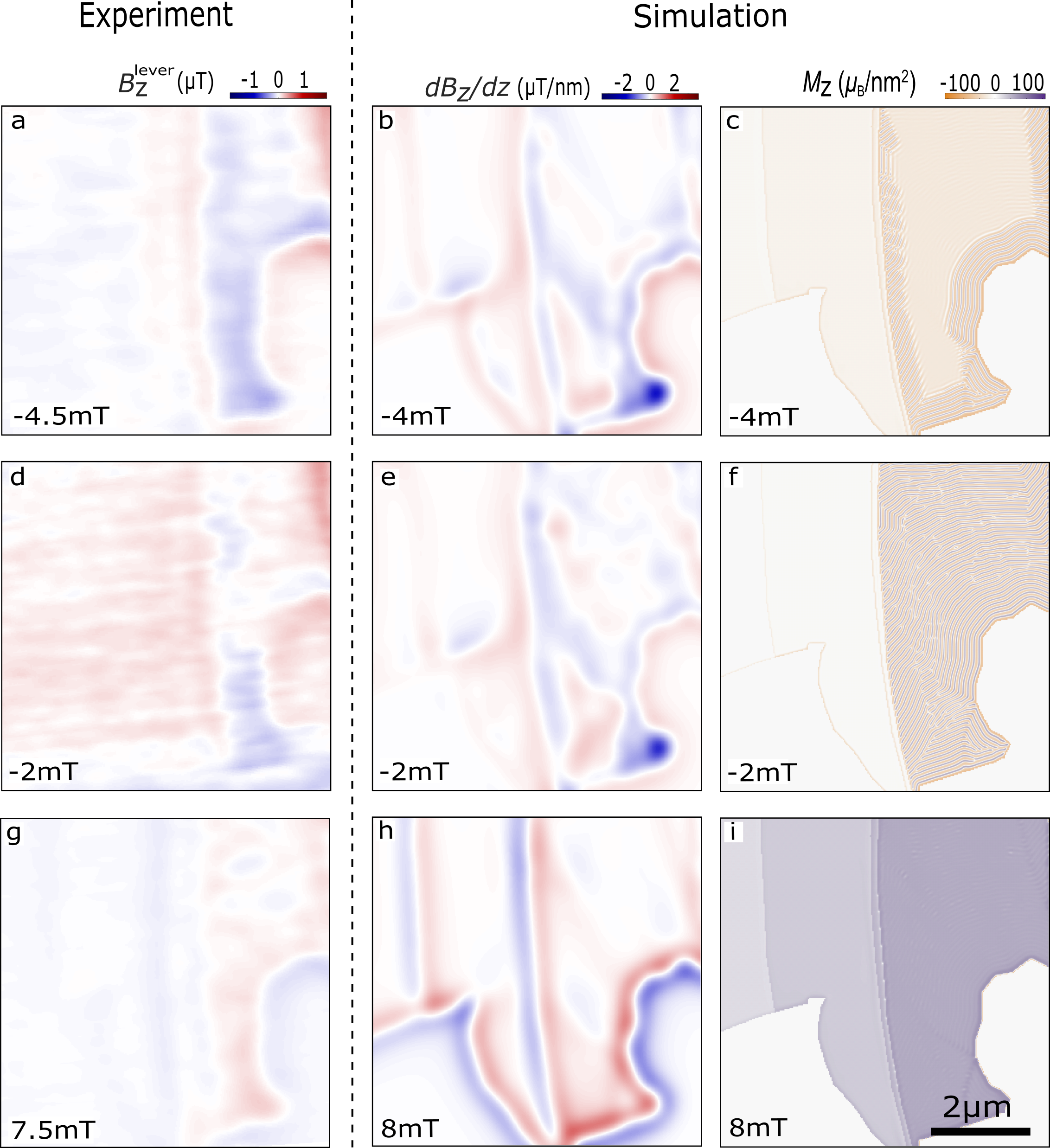}
\caption{Evolution of magnetic domains under increasing  $H_z$ for 3-5 layers. (a) $B_z^\text{lever}(x, y)$ measured at $\mu_0H_z = -4.5$~mT over the 3 to 5-layer-thick region of the sample indicated by green and purple dots in Figure~1a, together with (b) the corresponding simulation of $dB_z/dz(x, y)$ and (c) $M_z(x,y)$ under similar $\mu_0H_z$. (d-f) and (g-i) show again the measured $B_z^\text{lever}(x, y)$ together with corresponding simulations under externally applied $\mu_0H_z = -2$~mT and $\mu_0H_z = 7.5$~mT respectively.}\label{figS4_sim_5L}
\end{figure}

\begin{figure}[H]%
\renewcommand{\thefigure}{S11}
\centering
\includegraphics[width=1\textwidth]{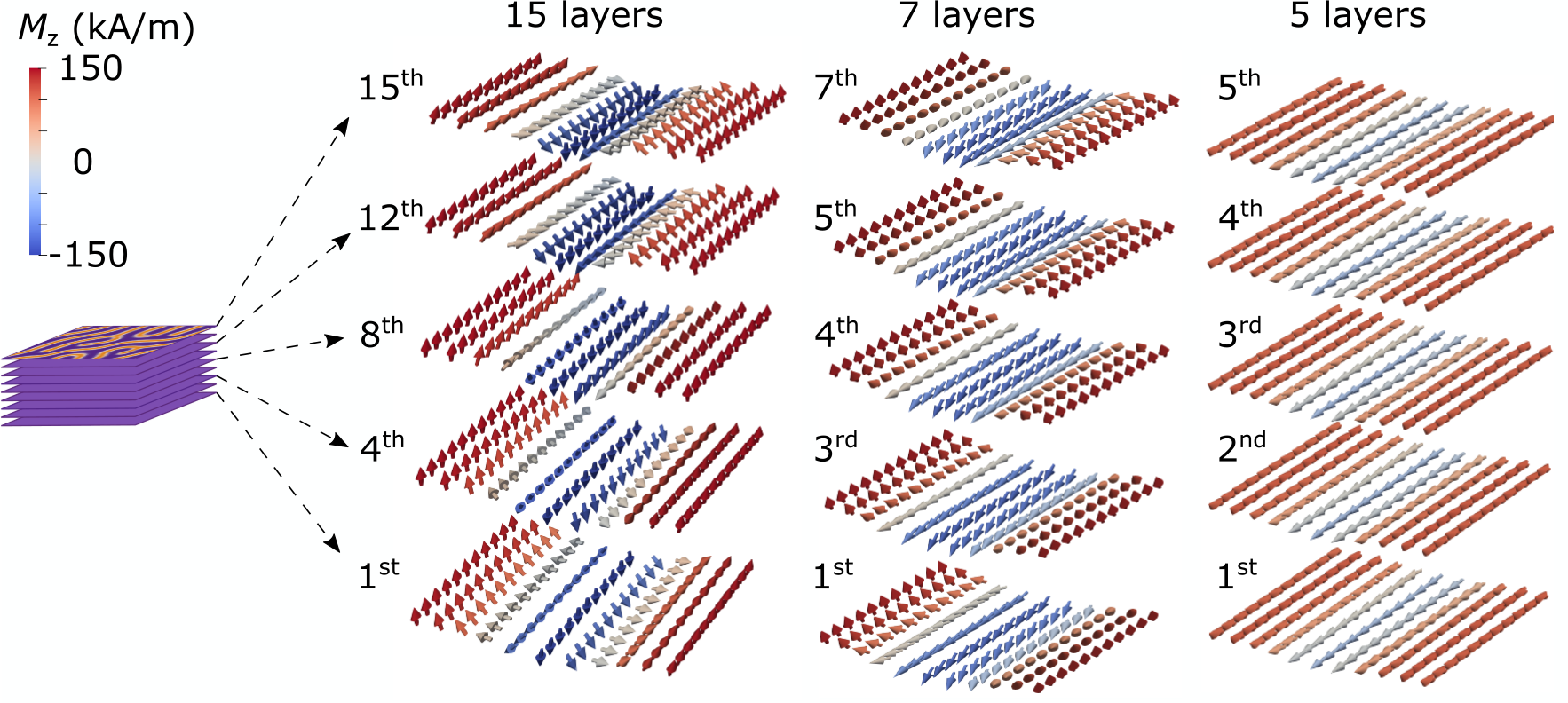}
\caption{Simulated spin texture of magnetic stripe domains for three thicknesses of the flake.}\label{figS5_spin}
\end{figure}

%
The simulated evolution of the magnetization texture under increasing $H_z$ for different areas of the measured CGT flake is presented in Fig~\ref{figS4_sim_15L}, ~\ref{figS4_sim_7L} and ~\ref{figS4_sim_5L}. The experimentally measured $B_z^\text{lever}(x, y)$ is compared with the simulated $dB_z(x, y)/dz$ and  $M_z(x,y)$ for selected  applied out-of-plane fields $H_z$.
Due to roughly one magnetically inactive layer found in the experiment, the simulation only considers the number of magnetically active layers.
%
\vspace{10mm}
As depicted in Figure~3, the simulated spin texture of CGT changes with the number of layers. 
Upon decreasing the number of layers, labyrinth domains evolve into parallel stripe domains. 
As depicted in Fig~\ref{figS5_spin} the underlying spin texture changes mostly due to the increasing effect of shape-anisotropy and to a smaller extent due to the a decrease in the magnetocrystalline anisotropy \(K_\text{u}\) and the intralayer exchange stiffness \(A_\text{ex}\).
The stripe domains within 15 layer CGT show a N\'eel-type domain wall on the top and bottom outer layers that progressively transitions into a Bloch-type wall towards the middle layer.
As thickness is decreased, the stripes have an increased in-plane magnetization component since the shape anisotropy also increases. Also, the stripes are topologically trivial as shown in Fig~\ref{figS5_spin} for 7 and 5 layers. However, there is still some small progressive change in the helicity of the domain walls from the top to the bottom layer.
Such effects are inferred from simulations, but cannot be directly imaged by our SSM technique due to a lack of spatial resolution and the fact that we measure $B_z$ rather than all vectorial components of the local magnetization. 
% ...
%
%
\vspace{20mm}
%
% ...
%
\begin{figure}[H]%
\renewcommand{\thefigure}{S12}
\centering
\includegraphics[width=1\textwidth]{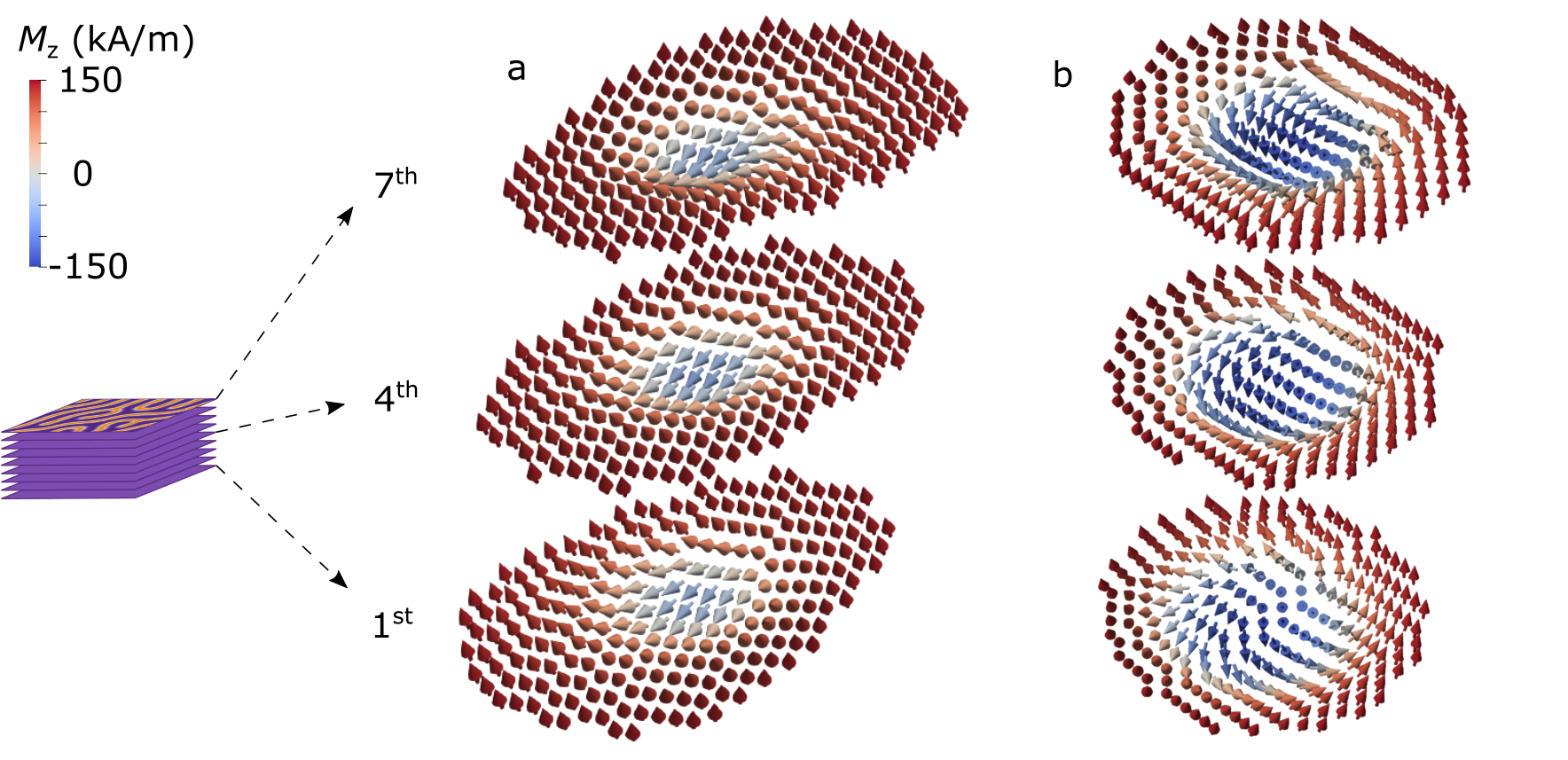}
\caption{Simulated spin texture of bubbles in 7-layer-thick CGT. (a) shows the spin texture of most bubbles found in this thickness, which are topologically trivial and (b) shows occassionally appearing bubbles with a skyrmionic texture.
}\label{figS5_7L_bubble}
\end{figure}